\PassOptionsToPackage{unicode}{hyperref}
\PassOptionsToPackage{hyphens}{url}

\documentclass[pdflatex,sn-nature]{nature/sn-jnl}

\usepackage{graphicx}
\usepackage{xcolor}
\usepackage{amsmath,amssymb}
\usepackage{booktabs,array}
\usepackage{tabularx} %
\usepackage{float} %
\usepackage[section]{placeins} %
\usepackage{multirow}

\begin{document}

\title{EPIC: Differentiable Energy-Force Consistent Uncertainty Quantification and Propagation for Machine-Learned Interatomic Potentials}

\author[1]{\fnm{Tianhao} \sur{Su}}
\equalcont{These authors contributed equally to this work.}
\author[1]{\fnm{Musen} \sur{Li}}
\equalcont{These authors contributed equally to this work.}
\author*[1,2,3]{\fnm{Shunbo} \sur{Hu}}\email{shunbohu@shu.edu.cn}
\author*[1]{\fnm{Tong-Yi} \sur{Zhang}}\email{zhangty@shu.edu.cn}
\affil[1]{\orgname{Materials Genome Institute, Shanghai University},
  \orgaddress{\city{Shanghai}, \postcode{200444}, \country{China}}}
\affil[2]{\orgname{Institute for the Conservation of Cultural Heritage, Institute for Quantum Science and Technology, Shanghai University},
  \orgaddress{\city{Shanghai}, \postcode{200444}, \country{China}}}
\affil[3]{\orgname{Key Laboratory of Silicate Cultural Heritage Conservation (Shanghai University), Ministry of Education},
  \orgaddress{\city{Shanghai}, \postcode{200444}, \country{China}}}

\abstract{\unboldmath

Machine-learned interatomic potentials (MLIPs) now rival density functional theory (DFT) in accuracy and are increasingly used to screen vast crystal databases, yet they can produce unreliable predictions for chemistries outside their training distribution without clear warning. Existing force-level evidential approaches do not inherently define a corresponding uncertainty model over the potential energy surface, limiting the consistent propagation of uncertainty to energy-derived observables. Here we introduce Energy-Propagated Intrinsic Confidence (EPIC), a general energy-centered evidential framework that derives atomic forces as exact gradients of the predicted mean energy, ensuring conservative force fields while providing analytical energy uncertainty, hierarchical atom-wise variance allocation, and energy-propagated $3\times 3$ force covariance tensors. Coupled to a MACE backbone and evaluated across the held-out Materials Project trajectory test partition ($79{,}020$ structures; $2{,}494{,}243$ atoms), EPIC achieves an energy mean absolute error of $0.0281$~eV/atom ($0.0292$~eV/atom across the unfiltered full set), top-5\% error detection AUROCs of $0.891$ for energy and $0.883$ for forces, and a calibrated 95\% force interval coverage of $96.07\%$, recalling $64.7\%$ of the top-5\% dangerous force-error structures within a $10\%$ DFT review budget ($\mathrm{AUROC} = 0.912$). Propagating uncertainty to equations of state, anisotropic force uncertainty ellipsoids, and harmonic phonon dispersions across $10{,}966$ candidate crystals reveals that the fraction classified as robustly dynamically stable rises from $11\%$ in the highest-uncertainty quartile to $53\%$ in the lowest-uncertainty quartile, while Monte Carlo imaginary-frequency excursions flag high-suspicion candidates where either the MLIP force prediction is locally inaccurate or the unrelaxed candidate structure itself is physically unreasonable. EPIC thus provides a physics-aware, chemistry-sensitive uncertainty diagnostic for high-throughput screening and autonomous materials discovery.

}
\maketitle

\section{Introduction}\label{sec:introduction}

Machine learning interatomic potentials (MLIPs) based on equivariant message passing neural networks, including NequIP~\citep{Batzner2022}, PaiNN~\citep{Schuett2021PaiNN}, and MACE~\citep{Batatia2022MACE}, routinely predict energies within a few meV/atom across diverse chemical systems. This computational efficiency enables molecular dynamics simulations at length and time scales previously inaccessible to first-principles methods~\citep{Ko2023,MISHIN2021116980}. Yet, when an MLIP encounters atomic configurations outside its training distribution, such as strained interfaces, exotic coordination environments, or unrepresented phases, it can produce plausible but severely erroneous predictions~\citep{MontesdeOcaZapiain2022,MISHIN2021116980}. Unlike density functional theory (DFT), which typically signals physical difficulties through self-consistent field (SCF) non-convergence, standard MLIPs fail silently. This absence of intrinsic reliability indicators restricts the deployment of MLIPs in materials discovery and high-throughput screening, where distinguishing reliable from unreliable predictions is essential~\citep{MISHIN2021116980,Riebesell2025}.

Uncertainty quantification (UQ) provides a principled mechanism to flag anomalous out-of-distribution (OOD) configurations and guide active learning campaigns to iteratively expand the training set~\citep{PhysRevMaterials.3.023804,Kulichenko2023}. Deep ensembles~\citep{Lakshminarayanan2017DeepEnsembles} remain the standard approach, offering well-characterized calibration at a linear multiplier of computational cost. While manageable for specific tasks, this overhead becomes prohibitive for large-scale simulations requiring millions of force evaluations. Alternative methods span various trade-offs: Monte Carlo (MC) dropout~\citep{Gal2016Dropout} provides lightweight but approximate Bayesian inference, whereas Gaussian process potentials like GAP~\citep{Bartok2010GAP} offer principled posterior uncertainty estimates. Because exact GP training scales cubically with dataset size, practical implementations must rely on sparse approximations to manage computational costs~\citep{pmlr-v5-titsias09a}. Evidential deep learning (EDL)~\citep{Amini2020DER} occupies a highly efficient position within this space. By placing a higher-order prior over the predictive distribution, EDL captures both aleatoric and epistemic uncertainty in a single model evaluation. Recent adaptations have integrated EDL into MLIPs. The eIP framework~\citep{Xu2026EvidentialIP} models each Cartesian force component independently using a scalar Normal-Inverse-Gamma (NIG) prior, whereas the subsequent e$^2$IP framework~\citep{Wang2026EquivariantEDLIP} introduces a multivariate Normal-Inverse-Wishart (NIW) formulation. Through rotation-equivariant parameterizations, e$^2$IP enables explicit modeling of anisotropic force uncertainties.

Despite their utility for fine-grained local force diagnostics and active learning, these evidential implementations formulate uncertainty strictly at the force level. Because their uncertainty distributions are parameterized directly over vector components, they do not inherently define a corresponding explicit scalar uncertainty model over the potential energy surface. In physical systems, however, atomic forces $\mathbf{F}_i = -\nabla_{\mathbf{R}_i} E(\mathbf{R})$ and local second-order curvatures $\Phi_{ij} = \partial^2 E(\mathbf{R}) / \partial \mathbf{R}_i \partial \mathbf{R}_j$ are not independent quantities, but are strictly locked to the scalar potential energy surface through differential constraints. While recent approaches such as Hessian Interatomic Potentials (HIP)~\citep{Burger2026HIP} bypass coordinate differentiation to predict second-order Hessians directly for computational speed, and force-level evidential models parameterize uncertainty without differentiating a scalar energy prior, detaching derivatives from the underlying energy surface sacrifices strict force conservativeness and provides no direct way to verify whether the local curvature of the learned energy landscape is physically reliable. An MLIP can exhibit small total-energy residuals and nearly vanishing forces at symmetric equilibrium sites while silently distorting the local potential-energy curvature that governs equations of state, phase stability rankings~\citep{Riebesell2025}, and harmonic phonon spectra~\citep{TOGO20151}. To preserve and exploit this fundamental differential relationship, we propose Energy-Propagated Intrinsic Confidence (EPIC), an energy-centered evidential framework universally applicable to autograd-based interatomic potentials. EPIC replaces the standard scalar energy readout with a hierarchical probabilistic evidential module, placing a Normal-Inverse-Gamma (NIG) distribution over the total potential energy alongside an atom-wise variance allocation head. The predictive mean energy acts as the scalar potential from which deterministic atomic forces are derived via automatic differentiation~\citep{Gao2020}, preserving mean-level energy-force consistency and ensuring conservative force fields by construction. Moreover, by coupling the learned energy-level epistemic evidence with last-layer feature-Jacobian covariance and differentiating the resulting conservative perturbed energy surfaces, EPIC propagates scalar energy uncertainty strictly along the physical derivative chain into closed-form $3\times 3$ atomic force covariance tensors and second-order curvature perturbations without requiring separate non-conservative heads.

We demonstrate the EPIC framework by integrating it into the MACE architecture. Functioning as a complement to local force-level models, EPIC provides a mathematically consistent route for uncertainty quantification in energy-centered atomistic workflows. Through systematic validation on multi-component materials from the MPtrj database~\citep{Deng2023}, we establish that the extracted epistemic uncertainty provides a robust relative ranking signal reflecting configurational data coverage, while post-hoc affine variance recalibration restores quantitative interval coverage across both energies and forces. By propagating this energy-level uncertainty to macroscopic physical observables, the framework yields diagnostic confidence bands for phonon dispersions~\citep{TOGO20151}, reveals explicit sampling biases in equation-of-state evolutions, and constructs real-space force uncertainty ellipsoids. Ultimately, by flagging high-suspicion dynamical instabilities and prioritizing uncertain candidates for first-principles verification in high-throughput screening~\citep{Tawfik2023,Riebesell2025}, EPIC establishes a practical, physics-aware foundation for reliable active learning and autonomous materials discovery.

\section{Results}\label{results}

\subsection{Predictive accuracy, force uncertainty propagation, and DFT review prioritization on the MPtrj dataset}\label{subsec:accuracy_calibration}

\begin{figure}[!htbp]
\centering
\includegraphics[width=\linewidth]{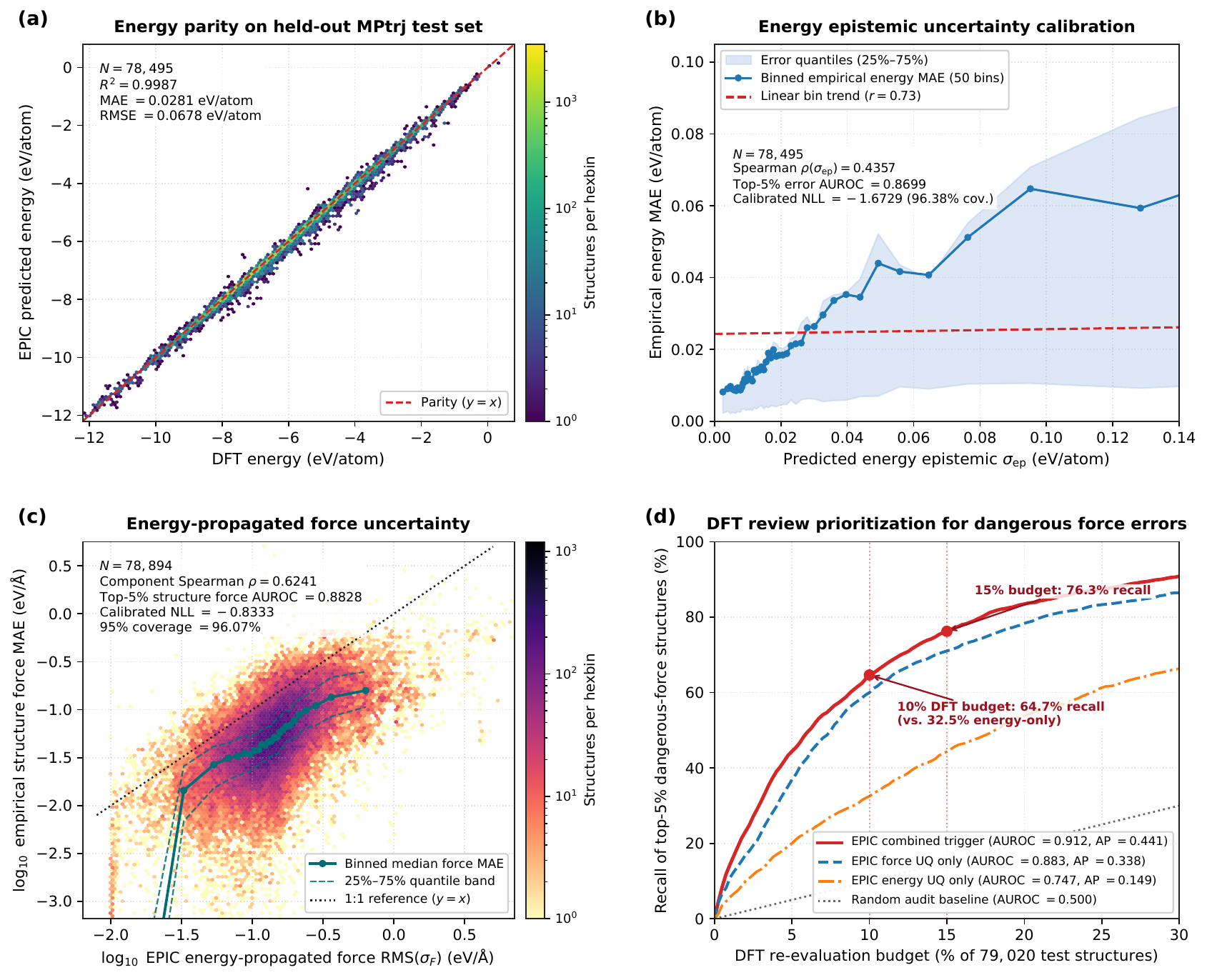}
\caption{Comprehensive evaluation of EPIC energy and energy-propagated force uncertainty on the held-out MPtrj test set. \textbf{(a)} Parity hexbin density plot of DFT versus EPIC-predicted energies across $78{,}495$ finite-variance test structures ($\mathrm{MAE} = 0.0281$~eV/atom, $\mathrm{RMSE} = 0.0678$~eV/atom, $R^2 = 0.9987$); dashed red line denotes $y = x$. \textbf{(b)} Energy uncertainty calibration curve across 50 equal-count bins sorted by epistemic $\sigma_{\mathrm{ep}}$, showing monotonic growth of mean $|\Delta E|$ (solid blue line) and interquartile range (shaded band), alongside finite-variance test structures ($N = 78{,}495$)  ranking and calibrated coverage metrics. \textbf{(c)} Log-log hexbin density of structure-level energy-propagated force uncertainty $\mathrm{RMS}(\sigma_F)$ versus actual DFT force MAE across all $79{,}020$ test structures ($7{,}482{,}729$ force components (The data and statistics shown in the figure are based on a total of 78,894 structures after excluding some entries with zero values.); component Spearman $\rho = 0.6241$, top-5\% AUROC $= 0.8333$, calibrated $\mathrm{NLL} = -0.6773$, 95\% coverage $= 96.07\%$). \textbf{(d)} Active-learning DFT review prioritization curves across $N = 79{,}020$ structures for recalling the top-5\% dangerous force-error configurations ($\mathrm{MAE}_F \ge P_{95} = 0.1515$~eV/\AA{}): the EPIC combined energy-force trigger achieves $\mathrm{AUROC} = 0.912$ ($\mathrm{AUPRC} = 0.441$) and recalls $64.7\%$ of dangerous structures within a $10\%$ DFT review budget, nearly doubling the recall of energy UQ alone ($32.5\%$).}
\label{fig:eval_uncertainty_analysis}
\end{figure}

To establish a baseline for predictive accuracy and evaluate the statistical reliability of the evidential uncertainty, we benchmarked MACE-EPIC on the held-out testing partition of the global MPtrj dataset~\citep{Deng2023}, which comprises $79{,}020$ crystal structures ($2{,}494{,}243$ atoms and $7{,}482{,}729$ Cartesian force components; dataset partitioning details are provided in Supplementary Note~3). Among the $79{,}020$ test structures, $78{,}495$ ($99.34\%$) satisfy the strict finite-variance condition $\alpha_G - 1 > 10^{-5}$ for the structure-level NIG shape parameter, while the remaining $525$ ($0.66\%$) boundary configurations lie near the $\alpha_G \to 1^+$ divergence threshold and are retained as the highest-uncertainty group in all full-set ($N = 79{,}020$) ranking, error-detection, and calibrated coverage evaluations (Supplementary Note~3). On the $78{,}495$ finite-variance test set, EPIC achieves a mean absolute error (MAE) of $0.0281$~eV/atom, a root-mean-square error (RMSE) of $0.0678$~eV/atom, and $R^2 = 0.9987$ for total energy predictions (Fig.~\ref{fig:eval_uncertainty_analysis}a); across all $79{,}020$ unfiltered test structures, the energy MAE is $0.0292$~eV/atom ($\mathrm{RMSE} = 0.0707$~eV/atom).

Beyond point-estimate accuracy, the NIG formulation natively decomposes total predictive energy uncertainty ($\sigma_{\mathrm{tot}} = \sqrt{\sigma_{\mathrm{ep}}^2 + \sigma_{\mathrm{al}}^2}$) into physically distinct epistemic (model-structural, $\sigma_{\mathrm{ep}}$) and aleatoric (data-intrinsic, $\sigma_{\mathrm{al}}$) contributions. Evaluating these raw components against the per-atom absolute energy prediction error ($|\Delta E|$) on the finite-variance set ($N = 78{,}495$) reveals positive monotonic correlations for both the epistemic ($\mathrm{median} = 16.61$~meV/atom, Spearman $\rho = 0.4357$; $\rho = 0.4431$ across the full $79{,}020$ test set) and aleatoric ($\mathrm{median} = 5.59$~meV/atom, Spearman $\rho = 0.4248$; $\rho = 0.4324$ across the full $79{,}020$ test set) components, with epistemic uncertainty consistently outperforming aleatoric uncertainty in error ranking (Table~\ref{tab:energy_uncertainty_coverage}). Crucially, because the epistemic-to-aleatoric ratio $\sigma_{\mathrm{ep}}/\sigma_{\mathrm{al}} = \nu_G^{-1/2}$ varies across configurations according to local evidence density $\nu_G$, combining both components into the total predictive uncertainty $\sigma_{\mathrm{tot}}$ ($\mathrm{median} = 18.12$~meV/atom) captures complementary model-driven and label-noise error sources (such as DFT relaxation trajectory and Hubbard-$U$ heterogeneity in MPtrj), boosting the Spearman rank correlation to $\rho = 0.4520$ on the finite-variance set and $\rho = 0.4591$ across the full $79{,}020$ test set (Table~\ref{tab:energy_uncertainty_coverage}).

\begin{table}[htbp]
  \centering
  \footnotesize
  \setlength{\tabcolsep}{4pt}
  \caption{Summary of uncalibrated energy uncertainty components on the finite-variance MPtrj test subset ($N = 78{,}495$, $\alpha_G - 1 > 10^{-5}$) and the full unfiltered test set ($N = 79{,}020$). Full-set calibrated energy and energy-propagated force uncertainty metrics are reported in Table~\ref{tab:full_79020_benchmark}.}
  \label{tab:energy_uncertainty_coverage}
  \begin{tabular}{@{}lcccc@{}}
    \toprule
    & \multicolumn{2}{c}{\textbf{Finite subset ($N = 78{,}495$)}}
    & \multicolumn{2}{c}{\textbf{Full test set ($N = 79{,}020$)}} \\
    \cmidrule(lr){2-3} \cmidrule(lr){4-5}
    \textbf{Uncertainty component}
    & \textbf{$\mathrm{median}(\sigma)$}
    & \textbf{Spearman $\rho$}
    & \textbf{$\mathrm{median}(\sigma)$}
    & \textbf{Spearman $\rho$} \\
    & \textbf{(meV/atom)}
    & \textbf{($\sigma, |\Delta E|$)}
    & \textbf{(meV/atom)}
    & \textbf{($\sigma, |\Delta E|$)} \\
    \midrule
    Epistemic ($\sigma_{\mathrm{ep}}$) & 16.61 & 0.4357 & 16.73 & 0.4431 \\
    Aleatoric ($\sigma_{\mathrm{al}}$) & 5.59  & 0.4248 & 5.65  & 0.4324 \\
    Total ($\sigma_{\mathrm{tot}} = \sqrt{\sigma_{\mathrm{ep}}^2 + \sigma_{\mathrm{al}}^2}$) & 18.12 & 0.4520 & 18.27 & 0.4591 \\
    \bottomrule
  \end{tabular}
\end{table}

To examine how raw evidential uncertainty relates to empirical errors across the full dynamic range, we sorted all test predictions by their epistemic uncertainty ($\sigma_{\mathrm{ep}}$) into 50 equal-count bins and evaluated the mean $|\Delta E|$ per bin (Fig.~\ref{fig:eval_uncertainty_analysis}b). The empirical error curve increases monotonically across all 50 bins, while the gap between raw $\sigma_{\mathrm{ep}}$ and observed error widens in the upper tail, indicating conservative (under-confident) raw dispersion on extreme out-of-distribution structures. Because evidential networks trained with regularized marginal likelihoods naturally inflate raw variance on unfamiliar configurations, raw $\sigma_{\mathrm{ep}}$ serves directly as a strictly positive monotonic ranking signal (achieving a top-5\% energy error detection AUROC of $0.8907$ across all $79{,}020$ structures), whereas quantitative Gaussian interval coverage is restored via post-hoc affine variance recalibration on the disjoint calibration split, $\sigma_{\mathrm{cal}} = \sqrt{\tau^2\sigma^2 + \sigma_0^2}$ (Supplementary Note~3). After fitting the two scalar calibration parameters ($\tau_E = 0.7278$, $\sigma_{0,E} = 0.0365$~eV/atom), EPIC achieves a calibrated energy negative log-likelihood (NLL) of $-1.6152$ and an empirical 95\% interval coverage of $96.00\%$ across all $79{,}020$ test structures (Table~\ref{tab:full_79020_benchmark}).

Crucially, because local force errors in complex crystals do not always manifest as large global energy residuals, relying on structure-level energy uncertainty alone is insufficient to flag localized atomic force failures (energy $\sigma_{\mathrm{ep}}$ versus structure force MAE yields Spearman $\rho = 0.2043$ and top-5\% AUROC $= 0.7471$; Supplementary Fig.~S1a). By coupling the damped structure-level energy evidence to the coordinate Jacobians of the last-layer invariant features (Methods, Eqs.~\eqref{eq:mc-energy-sampling}--\eqref{eq:closed-form-force-cov}), EPIC propagates energy-level confidence directly into anisotropic atomic force uncertainties. Across all $79{,}020$ test structures ($7{,}482{,}729$ force components), the resulting energy-propagated force uncertainty tightly tracks actual DFT force errors across four orders of magnitude (Fig.~\ref{fig:eval_uncertainty_analysis}c and Supplementary Fig.~S1b), achieving a component-level and structure-level Spearman correlation of $\rho = 0.6600$, a top-5\% error detection AUROC of $0.8831$ ($\mathrm{AUPRC} = 0.3385$), a calibrated component NLL of $-0.6773$ ($\tau_F = 48.2697$, $\sigma_{0,F} = 0.003374$~eV/\AA{}), and a 95\% interval coverage of $95.82\%$ (Table~\ref{tab:full_79020_benchmark}).

To further benchmark EPIC against established single-model and multi-model UQ baselines, we compared it on the full $79{,}020$-structure test set against analytic last-layer prediction rigidity (LLPR)~\citep{bigi2024prediction}, $256$-sample Monte Carlo LLPR (MC-LLPR), and a five-member deep ensemble (Ensemble-5), with complete metrics and wall-clock computational cost analyses provided in Supplementary Note~4 (Supplementary Tables~S3 and S4). For energy uncertainty quantification, EPIC achieves the highest Spearman correlation ($\rho = 0.4591$), top-5\% error AUROC ($0.8907$), and AUPRC ($0.4312$) at a $1.0\times$ single-model inference cost ($3.84$~ms/structure vs.\ $3.65$~ms/structure for LLPR), substantially outperforming both LLPR/MC-LLPR ($\rho = 0.2152$, $\mathrm{AUROC} = 0.6094$, $\mathrm{AUPRC} = 0.1131$) and Ensemble-5 ($\rho = 0.4119$, $\mathrm{AUROC} = 0.7222$, $\mathrm{AUPRC} = 0.2039$ at $5.0\times$ training and $4.9\times$ inference cost [$17.82$~ms/structure]). For force uncertainty quantification, Ensemble-5 provides the highest raw ranking correlation ($\rho = 0.7456$, $\mathrm{AUROC} = 0.8872$, $\mathrm{AUPRC} = 0.4444$) at five times the training and message-passing cost, whereas single-model EPIC outperforms both analytic LLPR ($\rho = 0.5952$, $\mathrm{AUROC} = 0.8311$, $\mathrm{AUPRC} = 0.2406$, $\mathrm{AtomNDCG} = 0.8955$) and MC-LLPR ($\rho = 0.5984$, $\mathrm{AUROC} = 0.8326$, $\mathrm{AUPRC} = 0.2490$, $\mathrm{AtomNDCG} = 0.8957$) across every metric ($\rho = 0.6600$, $\mathrm{AUROC} = 0.8831$, $\mathrm{AUPRC} = 0.3385$, $\mathrm{AtomNDCG} = 0.8973$) and matches or exceeds the ensemble's top-5\% error discrimination ($\mathrm{AUROC} = 0.8831$ at the component and structure levels, and $0.9116$ with the combined energy-force trigger) while achieving the lowest calibrated negative log-likelihood ($\mathrm{NLL} = -0.6773$ vs.\ $-0.2672$ for LLPR, $-0.3466$ for MC-LLPR, and $-0.3339$ for Ensemble-5) and accurate 95\% interval coverage ($95.82\%$).

Finally, we evaluated the practical utility of EPIC as an automated gatekeeper for prioritizing high-risk structures for first-principles DFT verification (Fig.~\ref{fig:eval_uncertainty_analysis}d and Table~\ref{tab:full_79020_benchmark}). Defining the top 5\% of test structures with the largest DFT force errors ($\mathrm{MAE}_F \ge P_{95} = 0.2015$~eV/\AA{}, $3{,}951$ structures) as dangerous force-prediction failures, we constructed a composite score by combining EPIC's energy epistemic uncertainty, energy-propagated force uncertainty, and maximum predicted force magnitude. The percentile ranks of force uncertainty, energy epistemic uncertainty, and maximum predicted force magnitude were assigned weights of 0.5, 0.2, and 0.3, respectively. Ranking candidate structures by this score achieved an AUROC of 0.9116 and an AUPRC of 0.4415 for identifying dangerous force-prediction failures. Within a $10\%$ DFT review budget ($7{,}902$ queried structures), the EPIC trigger captures $64.7\%$ of all dangerous force failures---nearly $6.5\times$ the random-sampling baseline ($10.0\%$) and nearly double the recall obtained using energy uncertainty alone ($32.5\%$). Expanding the review budget to $15\%$ and $20\%$ recalls $76.3\%$ and $83.5\%$ of all dangerous structures, respectively, demonstrating that energy-propagated force uncertainty is essential for reliable single-model active learning and error screening. Extended analyses of the NIG posterior phase space, aleatoric--epistemic decoupling across driving forces, within-structure atomic error localization, multi-model Pareto frontiers, three-stage curriculum training dynamics, and system-size scaling invariance ($N_{\mathrm{atoms}} \in [1, 336]$) are presented in Supplementary Notes~6 and 11 (Supplementary Figs.~S6 and S11).

\begin{table}[htbp]
  \centering
  \footnotesize
  \setlength{\tabcolsep}{3pt}
  \caption{Full-set ($N = 79{,}020$ structures; $2{,}494{,}243$ atoms; $7{,}482{,}729$ force components) uncertainty quantification, post-hoc calibration, and DFT review prioritization metrics of MACE-EPIC on the held-out MPtrj test partition. Subset ``finite'' ($N=78{,}495$) satisfies $\alpha_G - 1 > 10^{-5}$; ``full'' ($N=79{,}020$) retains all $525$ ($0.66\%$) boundary structures ($\alpha_G - 1 \le 10^{-5}$). For $E/N$, Spearman $\rho$ reports total predictive uncertainty $\sigma_{\mathrm{tot}}$ with epistemic $\sigma_{\mathrm{ep}}$ in parentheses (Table~\ref{tab:energy_uncertainty_coverage}). Predictive energy accuracy is $\mathrm{MAE}=0.0281$~eV/atom ($\mathrm{RMSE}=0.0678$~eV/atom) on the finite-variance set and $0.0292$~eV/atom ($0.0707$~eV/atom) on the full set. Post-hoc affine calibration parameters: $\tau_E=1.007753$, $\sigma_{0,E}=0.02228$~eV/atom; $\tau_F=47.4231$, $\sigma_{0,F}=0.01024$~eV/\AA{}.}
  \label{tab:full_79020_benchmark}
  \begin{tabular}{@{}lcccc@{}}
    \toprule
    \textbf{Observable / Subset} & \textbf{Evaluation} & \textbf{Spearman} & \textbf{Top-5\% Error} & \textbf{Cal.\ NLL /} \\
    (eV/atom or eV/\AA{}) & \textbf{Scope ($N$)} & $\boldsymbol{\rho}$ & \textbf{AUROC / AUPRC} & \textbf{95\% Cov.} \\
    \midrule
    $E/N$, finite subset & $78{,}495$ struct. & $0.4520$ ($0.4357$) & $0.8817$ / $0.3505$ & $-1.6729$ / $96.38\%$ \\
    $E/N$, full test set & $79{,}020$ struct. & $0.4591$ ($0.4431$) & $0.8907$ / $0.3987$ & $-1.6253$ / $96.40\%$ \\
    $F_{i\alpha}$, component-level & $7.48\times 10^6$ comp. & $0.6242$ & $0.8649$ / $0.2803$ & $-0.8334$ / $96.07\%$ \\
    Struct.\ $\mathrm{MAE}_F$ & $79{,}020$ struct. & $0.6600$ & $0.8831$ / $0.3385$ & --- \\
    \midrule
    \textbf{DFT review trigger} & \textbf{Recall} & \textbf{Recall} & \textbf{Recall} & \textbf{Recall @ 20\%} \\
    ($\mathrm{MAE}_F \ge 0.2015$~eV/\AA{}) & \textbf{@ 5\%} & \textbf{@ 10\%} & \textbf{@ 15\%} & \textbf{(AUROC)} \\
    \midrule
    Energy $\sigma_{\mathrm{ep},G}/N$ alone & $20.0\%$ & $32.5\%$ & $44.4\%$ & $53.9\%$ ($0.7471$) \\
    Force $\mathrm{RMS}(\sigma_F)$ alone & $36.9\%$ & $60.0\%$ & $71.0\%$ & $78.3\%$ ($0.8831$) \\
    \textbf{EPIC combined trigger} & $\mathbf{44.4\%}$ & $\mathbf{64.7\%}$ & $\mathbf{76.3\%}$ & $\mathbf{83.5\%}$ ($\mathbf{0.9116}$) \\
    \bottomrule
  \end{tabular}
\end{table}

\subsection{Energy-volume topography exposes sampling biases}\label{subsec:thermo_self_consistency}

\begin{figure}[!htbp]
\centering
\includegraphics[width=\linewidth]{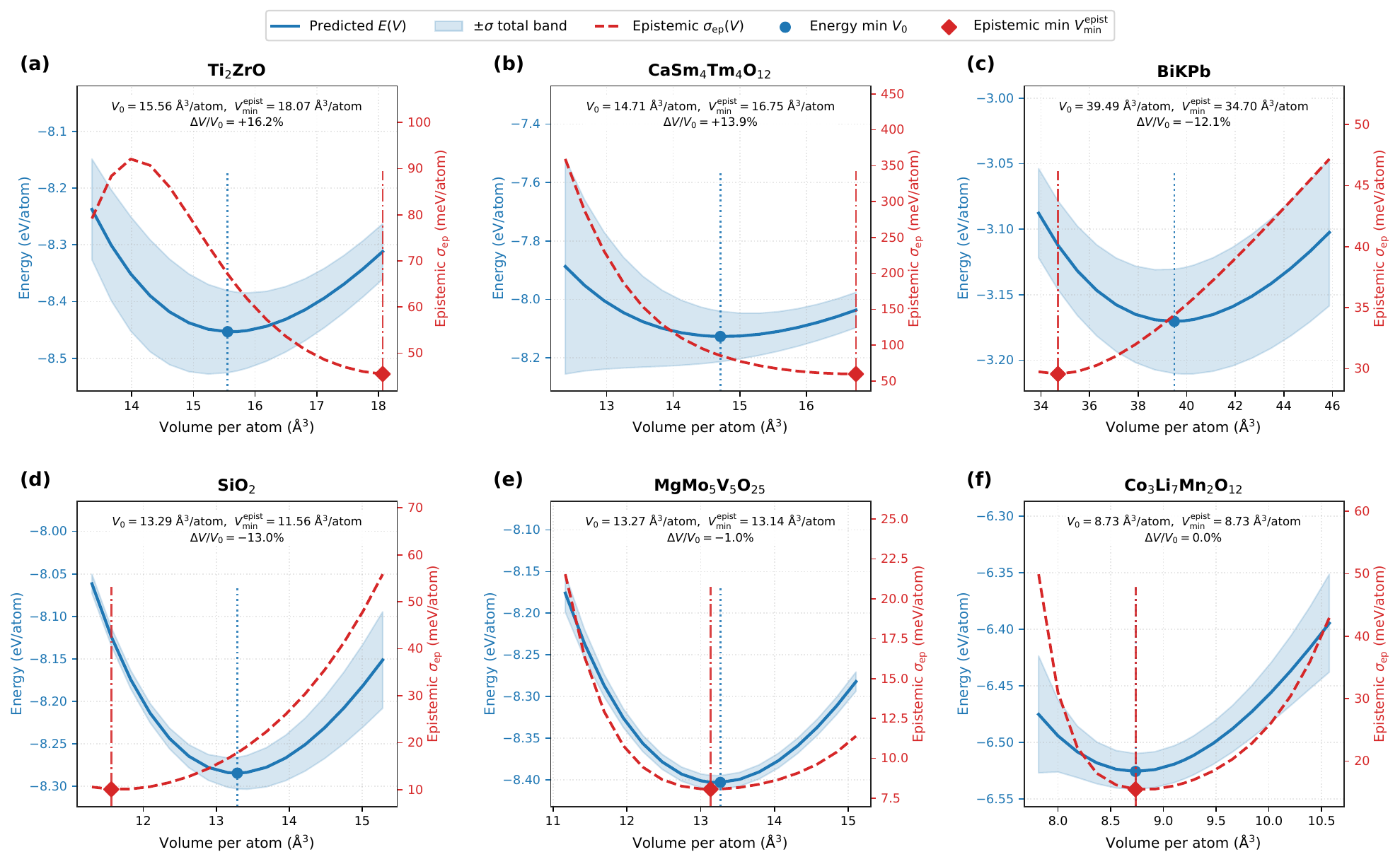}
\caption{Energy-volume curves and epistemic uncertainty distributions for six representative crystal structures: \textbf{(a)} $\mathrm{Ti_2ZrO}$, \textbf{(b)} $\mathrm{CaSm_4Tm_4O_{12}}$, \textbf{(c)} $\mathrm{BiKPb}$, \textbf{(d)} $\mathrm{SiO_2}$, \textbf{(e)} $\mathrm{MgMo_5V_5O_{25}}$, and \textbf{(f)} $\mathrm{Co_3Li_7Mn_2O_{12}}$. Solid blue lines denote the predicted mean energy (eV/atom), and light blue shaded regions represent the $\pm \sigma$ total uncertainty bands. Dashed red lines (right axis) indicate the epistemic uncertainty $\sigma_{\mathrm{ep}}$ (meV/atom). Vertical blue dotted lines and circles mark the predicted equilibrium volume $V_0$, whereas vertical red dash-dotted lines and diamonds mark the volume of minimum epistemic uncertainty $V_{\min}^{\mathrm{epist}}$. Inset boxes report $V_0$, $V_{\min}^{\mathrm{epist}}$, and the relative volume shift $\Delta V/V_0 = (V_{\min}^{\mathrm{epist}} - V_0)/V_0$.}
\label{fig:ev_combined_figure}
\end{figure}

To examine the local thermodynamic self-consistency of the framework beyond ensemble-level statistics, we probed the behavior of individual crystal configurations subjected to extreme lattice deformations. Across a benchmark set of $54$ structurally and chemically diverse crystals from the held-out test set, each structure was subjected to continuous, isotropic volume scaling from $0.85 V_0$ to $1.15 V_0$ (linear lattice scaling $s \in [0.947, 1.048]$), where $V_0$ corresponds to the thermodynamic equilibrium volume minimizing the predicted energy curve. Figure~\ref{fig:ev_combined_figure} highlights six representative materials spanning intermetallics, oxides, and silicate frameworks.

As displayed in Figure~\ref{fig:ev_combined_figure}, the predicted mean energies (solid blue lines) across both the compressive and tensile regimes map to smooth curves that conform closely to the classical Birch--Murnaghan equation of state~\citep{Birch1947,Murnaghan1944}. This seamless thermodynamic behavior highlights a core architectural strength of the EPIC framework. Because atomic forces are evaluated as the exact analytical gradients of the predicted mean energy, the underlying force field remains strictly conservative. Consequently, thermodynamic self-consistency is natively preserved even under severe volume deformation, circumventing the discontinuous force integration artifacts that can affect non-conservative models.

Tracking the epistemic uncertainty (red dashed lines, Fig.~\ref{fig:ev_combined_figure}) alongside the equation of state reveals an important structural offset. The global minimum of the epistemic uncertainty ($V_{\min}^{\mathrm{epist}}$, red diamonds), which corresponds to the state of highest model confidence, does not necessarily coincide with the physical equilibrium volume ($V_0$, blue circles). As quantified in Table~\ref{tab:ev_uncertainty_data}, $V_{\min}^{\mathrm{epist}}$ can deviate from $V_0$ by up to $+16.2\%$ in expansion or $-13.0\%$ in compression. Across the complete $54$-structure deformation benchmark ($886$ deformed configurations), $13$ structures ($24.1\%$) exhibit an epistemic uncertainty minimum aligned with $V_0$ within $|\Delta V/V_0| \le 3\%$, whereas $25$ structures ($46.3\%$) exhibit left-skewed minima shifted toward compressed volumes ($\Delta V/V_0 < -3\%$) and $16$ structures ($29.6\%$) exhibit right-skewed minima shifted toward expanded volumes ($\Delta V/V_0 > +3\%$). This separation illustrates a fundamental distinction between physical equilibrium and data-driven confidence: $V_0$ is governed by the underlying interatomic Hamiltonian, whereas $\sigma_{\mathrm{ep}}(V)$ is governed by the local configurational density of the training set along the volumetric coordinate. Specifically, the uncertainty profiles fall into three regimes: (i) U-shaped landscapes where the uncertainty minimum aligns closely with $V_0$ (e.g., $\mathrm{MgMo_5V_5O_{25}}$ with $\Delta V/V_0 = -1.0\%$ and $\mathrm{Co_3Li_7Mn_2O_{12}}$ with $\Delta V/V_0 = 0.0\%$), (ii) right-skewed profiles where the model exhibits maximum confidence in expanded states (e.g., $\mathrm{Ti_2ZrO}$ at $+16.2\%$ and $\mathrm{CaSm_4Tm_4O_{12}}$ at $+13.9\%$), and (iii) left-skewed profiles showing minimized uncertainty under compression (e.g., $\mathrm{BiKPb}$ at $-12.1\%$ and $\mathrm{SiO_2}$ at $-13.0\%$).

\begin{table}[htbp]
    \centering
    \footnotesize
    \setlength{\tabcolsep}{8pt}
    \caption{Energy-volume characteristics and locations of minimum epistemic uncertainty for the six representative crystal structures shown in Fig.~\ref{fig:ev_combined_figure}. The relative volume shift is defined as $\Delta V/V_0 = (V_{\min}^{\mathrm{epist}}-V_0)/V_0 \times 100\%$.}
    \label{tab:ev_uncertainty_data}
    \begin{tabular}{lcccc}
        \toprule
        \textbf{Formula}
        & $\boldsymbol{E_{\mathrm{eq}}/N}$
        & $\boldsymbol{V_0/N}$
        & $\boldsymbol{V_{\min}^{\mathrm{epist}}/N}$
        & $\boldsymbol{\Delta V/V_0}$ \\
        & (eV/atom)
        & (\AA$^3$/atom)
        & (\AA$^3$/atom)
        & (\%) \\
        \midrule
        $\mathrm{Ti_2ZrO}$
        & $-8.453$ & 15.56 & 18.07 & $+16.2$ \\
        $\mathrm{CaSm_4Tm_4O_{12}}$
        & $-8.126$ & 14.71 & 16.75 & $+13.9$ \\
        $\mathrm{BiKPb}$
        & $-3.170$ & 39.49 & 34.70 & $-12.1$ \\
        $\mathrm{SiO_2}$
        & $-8.285$ & 13.29 & 11.56 & $-13.0$ \\
        $\mathrm{MgMo_5V_5O_{25}}$
        & $-8.403$ & 13.27 & 13.14 & $-1.0$ \\
        $\mathrm{Co_3Li_7Mn_2O_{12}}$
        & $-6.526$ & 8.73 & 8.73 & $0.0$ \\
        \bottomrule
    \end{tabular}
\end{table}

These asymmetric uncertainty profiles reflect historical sampling biases along volumetric deformation coordinates in the training database, such as PBE-GGA volume overestimation relative to experimental initial structures or high-pressure polymorph trajectories. Within an active learning workflow, the directional shift $\Delta V/V_0$ provides a quantitative physical diagnostic to guide targeted DFT single-point acquisitions under specific strain states, systematically flattening the epistemic uncertainty landscape across the equation of state (see Supplementary Note~9 and Supplementary Fig.~S9 for the full $54$-crystal EOS curvature and tensile--compressive steepness analysis).

\subsection{Force uncertainty ellipsoids in real space}\label{subsec:uncertainty_ellipsoids}

A major theoretical advantage of the EPIC framework is its ability to recover the full tensorial anisotropy of atomic force uncertainties without requiring complex multi-dimensional matrix-variate priors on vector heads. Because the energy-level epistemic evidence modulates a differentiable perturbation of the pooled invariant features (Methods, Eqs.~\eqref{eq:mc-energy-sampling}--\eqref{eq:closed-form-force-cov}), differentiating with respect to the Cartesian coordinates $\mathbf{R}_i$ yields a complete, symmetric positive semi-definite $3 \times 3$ force covariance matrix $\Sigma_{\mathbf{F}_i}(\mathbf{R}) = \tau_F^2\,c_0^2\,J_{h,i}(\mathbf{R})\Sigma_F J_{h,i}(\mathbf{R})^{\top}$ for each atom $i$. Diagonalizing $\Sigma_{\mathbf{F}_i}(\mathbf{R})$ produces three mutually orthogonal principal axes with standard-deviation semi-axis lengths $\sqrt{\lambda_{i,1}} \le \sqrt{\lambda_{i,2}} \le \sqrt{\lambda_{i,3}}$, which geometrically define a force uncertainty ellipsoid in real space closely analogous to crystallographic anisotropic displacement parameters (ADPs).

\begin{figure}[!htbp]
\centering
\includegraphics[width=\linewidth]{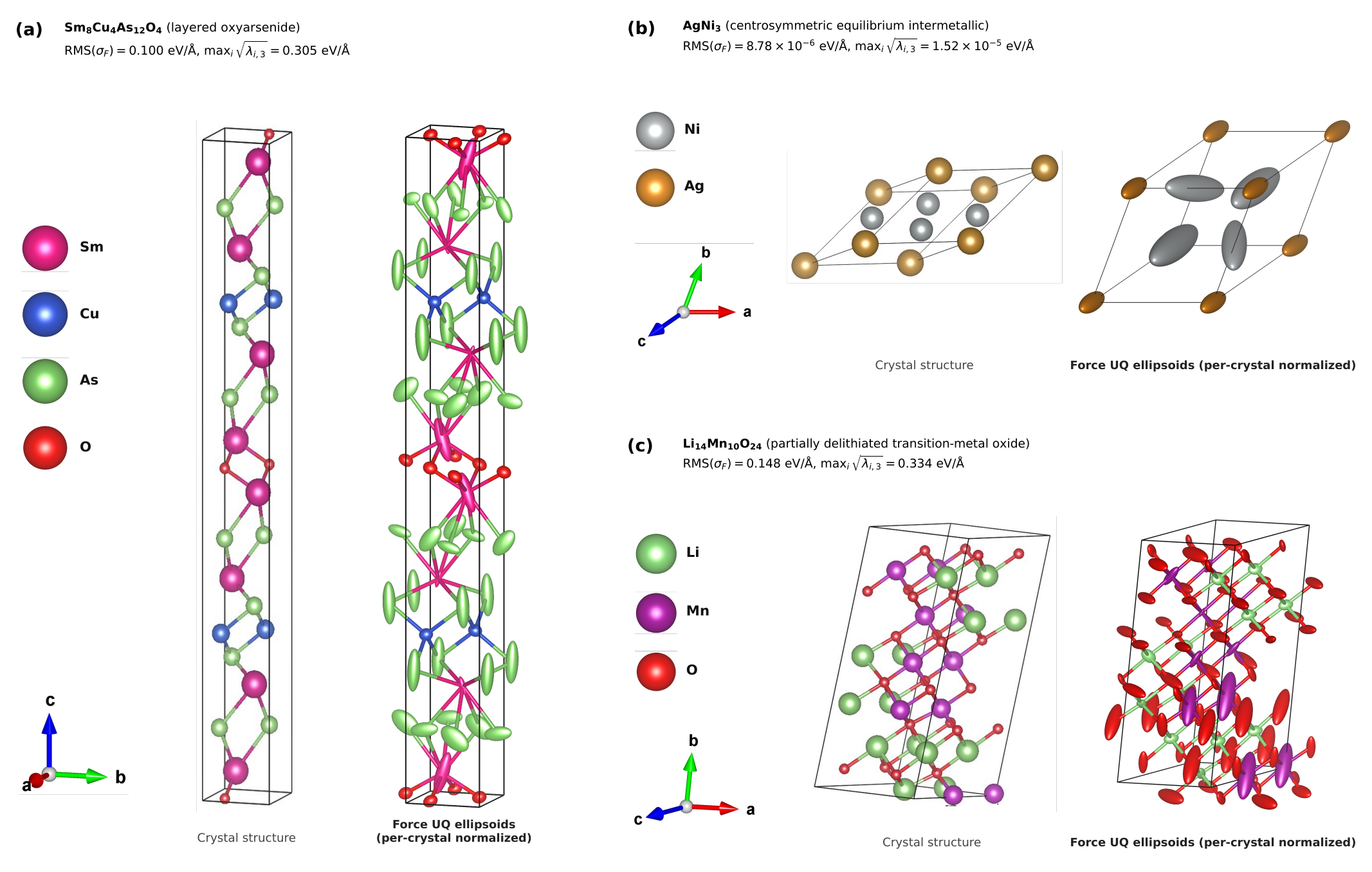}
\caption{Real-space visualization of atomic force epistemic uncertainty ellipsoids derived from the EPIC energy-propagated $3\times 3$ force covariance $\Sigma_{\mathbf{F}_i}(\mathbf{R})$, rendered in VESTA alongside the reference crystal structures (ellipsoids normalized per structure for visual clarity of relative site anisotropy; exact physical magnitudes are reported in each panel header). \textbf{(a)} Layered oxyarsenide $\mathrm{Sm_8Cu_4As_{12}O_4}$, exhibiting moderate overall uncertainty ($\mathrm{RMS}(\sigma_F) = 0.084$~eV/\AA{}) with localized directional elongation on the $\mathrm{Sm}$ sites ($\max_i\sqrt{\lambda_{i,3}} = 0.311$~eV/\AA{}). \textbf{(b)} Close-packed centrosymmetric intermetallic $\mathrm{AgNi_3}$, where crystallographic $O_h$ inversion symmetry strongly suppresses epistemic force variance ($\mathrm{RMS}(\sigma_F) = 8.9\times 10^{-6}$~eV/\AA{}, $\max_i\sqrt{\lambda_{i,3}} = 1.5\times 10^{-5}$~eV/\AA{}) and aligns the residual principal axes with crystallographic bonding planes. \textbf{(c)} Partially delithiated transition-metal oxide $\mathrm{Li_{14}Mn_{10}O_{24}}$, where oxygen anions (red) and redox-active $\mathrm{Mn}$ sites exhibit substantially larger and more diffuse uncertainty ellipsoids ($\mathrm{RMS}(\sigma_F) = 0.131$~eV/\AA{}, $\max_i\sqrt{\lambda_{i,3}} = 0.340$~eV/\AA{}) than the alkali sites.}
\label{fig:uncertainty_ellipsoids}
\end{figure}

As illustrated in Figure~\ref{fig:uncertainty_ellipsoids} (where ellipsoid glyphs are normalized within each crystal to visualize site-to-site directional anisotropy while exact physical magnitudes in eV/\AA{} are annotated in each header), the resulting ellipsoids resolve both structural anisotropy and site-specific chemical sensitivity. In the layered oxyarsenide $\mathrm{Sm_8Cu_4As_{12}O_4}$ (Fig.~\ref{fig:uncertainty_ellipsoids}a), the structure-level root-mean-square force uncertainty is moderate ($\mathrm{RMS}(\sigma_F) = 0.084$~eV/\AA{}), with compact ellipsoids on the covalent $\mathrm{CuAs_4}$ and $\mathrm{Sm_4O}$ layers and selective directional elongation at the rare-earth $\mathrm{Sm}$ sites ($\max_i\sqrt{\lambda_{i,3}} = 0.311$~eV/\AA{}). By contrast, the close-packed intermetallic $\mathrm{AgNi_3}$ (Fig.~\ref{fig:uncertainty_ellipsoids}b) occupies centrosymmetric $O_h$ ($1a/3c$) Wyckoff sites where inversion symmetry forces first-order invariant feature gradients to vanish, suppressing force uncertainty by four orders of magnitude ($\mathrm{RMS}(\sigma_F) = 8.9\times 10^{-6}$~eV/\AA{}, $\max_i\sqrt{\lambda_{i,3}} = 1.5\times 10^{-5}$~eV/\AA{}). Conversely, in the partially delithiated transition-metal oxide $\mathrm{Li_{14}Mn_{10}O_{24}}$ (Fig.~\ref{fig:uncertainty_ellipsoids}c), the force uncertainty rises markedly to $\mathrm{RMS}(\sigma_F) = 0.131$~eV/\AA{} ($\max_i\sqrt{\lambda_{i,3}} = 0.340$~eV/\AA{}), with large, anisotropic ellipsoids concentrated on the oxygen sublattice (red) and distorted $\mathrm{MnO_6}$ octahedra, while $\mathrm{Li}$ sites remain comparatively compact. This site-resolved contrast becomes even more pronounced in out-of-distribution complex oxides from the high-error tail of the test set (Supplementary Fig.~S2), such as $\mathrm{HoMoO_5}$ ($\mathrm{MAE}_F = 0.348$~eV/\AA{}, $\mathrm{RMS}(\sigma_F) = 0.223$~eV/\AA{}, $\max_i\sqrt{\lambda_{i,3}} = 0.414$~eV/\AA{}), $\mathrm{AgNiO_2}$ ($\mathrm{MAE}_F = 0.562$~eV/\AA{}, $\mathrm{RMS}(\sigma_F) = 0.133$~eV/\AA{}, $\max_i\sqrt{\lambda_{i,3}} = 0.460$~eV/\AA{}), and $\mathrm{Co_3Li_7Mn_2O_{12}}$ ($\mathrm{MAE}_F = 0.543$~eV/\AA{}, $\mathrm{RMS}(\sigma_F) = 0.143$~eV/\AA{}, $\max_i\sqrt{\lambda_{i,3}} = 0.423$~eV/\AA{}), visually pinpointing the exact atomic sublattices where local message-passing descriptors struggle to resolve complex charge-transfer and Jahn--Teller distortion environments. Furthermore, because $\mathbf{F}_i = -\nabla_{\mathbf{R}_i} E_G$ is the gradient of an $E(3)$-invariant scalar potential, the rank and triaxiality of $\Sigma_{\mathbf{F}_i}(\mathbf{R})$ rigorously obey crystallographic Wyckoff site-symmetry selection rules (Neumann's principle; Supplementary Note~8 and Supplementary Fig.~S8).

\subsection{Uncertainty-aware phonon screening and chemistry-dependent stability breakdown}\label{subsec:phonon_screening}

\begin{figure}[!htbp]
\centering
\includegraphics[width=\linewidth]{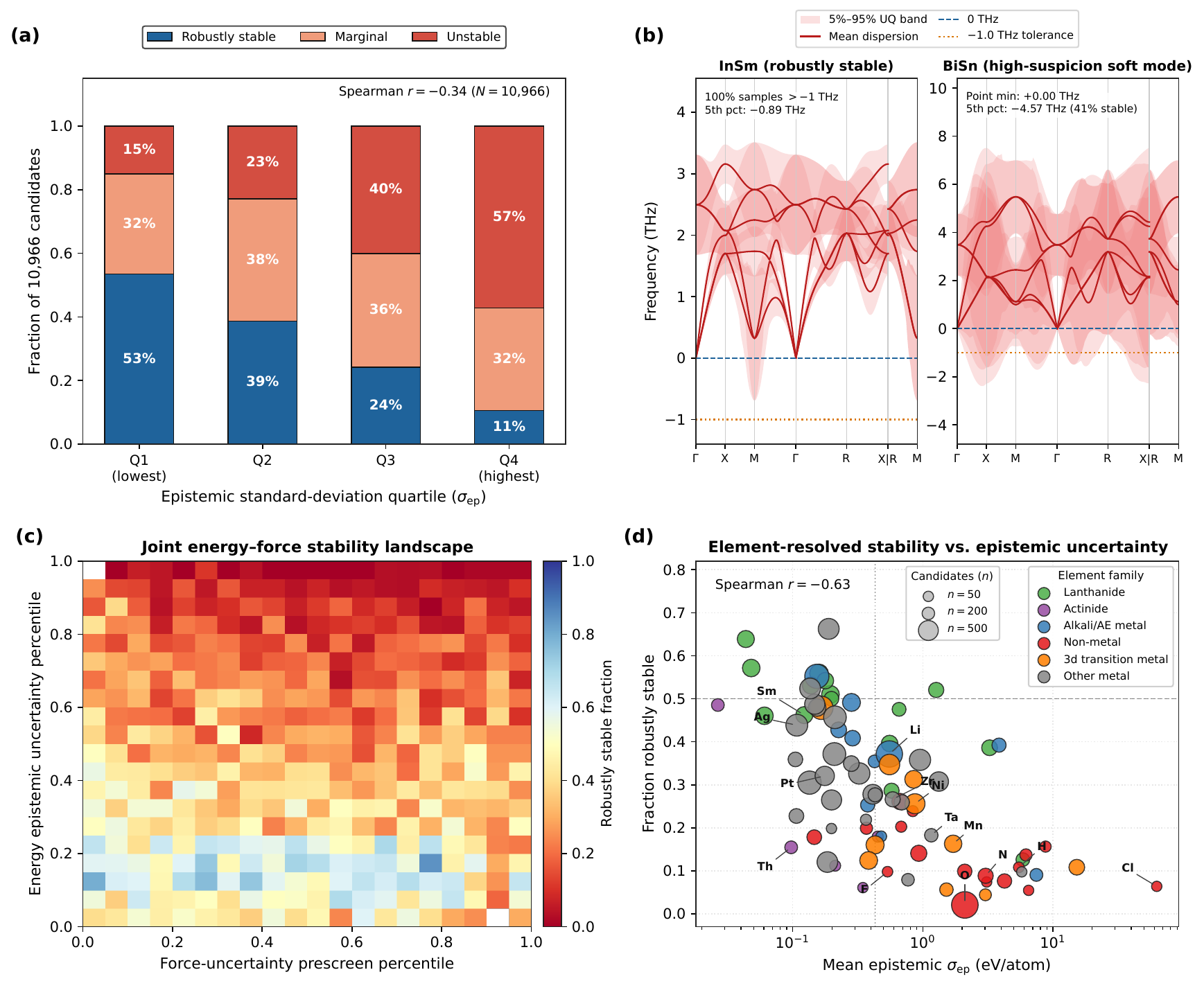}
\caption{Uncertainty-aware harmonic phonon stability screening across $10{,}966$ candidate crystals. \textbf{(a)} Fraction of robustly stable ($31.7\%$ overall), marginal ($34.5\%$), and unstable ($33.8\%$) structures within each quartile of epistemic uncertainty (Q1: lowest, Q4: highest; Spearman $r = -0.34$). \textbf{(b)} Side-by-side phonon dispersion comparison illustrating how energy-propagated force uncertainty separates genuinely stable candidates from high-suspicion false-stable predictions: in $\mathrm{InSm}$ (left, robustly stable), both the point prediction ($\omega_{\min}^{\mathrm{pt}} = 0.00$~THz) and the 5th-percentile Monte Carlo minimum ($\omega_{\min,q05} = -0.89$~THz, $P_{\mathrm{stable}} = 100\%$) remain above the $-1.0$~THz stability threshold, whereas in $\mathrm{BiSn}$ (right, marginal / high-suspicion soft mode), the deterministic point prediction appears dynamically stable ($\omega_{\min}^{\mathrm{pt}} = 0.00$~THz) but local force uncertainty drives severe imaginary-frequency excursions down to $\omega_{\min,q05} = -4.57$~THz ($P_{\mathrm{stable}} = 40.6\%$). Solid red lines denote point-predicted dispersions; shaded bands denote $\pm 1.645\sigma$ ($90\%$) Monte Carlo confidence intervals. \textbf{(c)} Two-dimensional binned map of the robustly stable fraction as a function of force-uncertainty prescreen percentile and energy epistemic uncertainty percentile; bins with fewer than five structures are masked. \textbf{(d)} Per-element robustly stable fraction versus mean epistemic uncertainty across elements ($n \ge 15$ structures, $r = -0.63$); bubble area is proportional to occurrence count.}
\label{deep_fig_combined}
\end{figure}

To evaluate how energy-propagated force uncertainty informs high-throughput dynamical stability screening, we computed finite-displacement harmonic phonon dispersions and $K = 32$ conservative Monte Carlo force-constant realizations across $10{,}966$ (For details on the origin of the structure, see Supplementary Note~3.) candidate crystal structures (Methods, Section~\ref{uncertainty-propagation-to-downstream-observables}). Combining the deterministic point-predicted minimum phonon frequency ($\omega_{\min}^{\mathrm{pt}}$) with the 5th percentile ($\omega_{\min,q05}$) and stable fraction ($P_{\mathrm{stable}}$) of the Monte Carlo dispersion ensemble under a $-1.0$~THz tolerance classifies the $10{,}966$ candidates into $3{,}478$ ($31.7\%$) \textbf{robustly stable}, $3{,}782$ ($34.5\%$) \textbf{marginal}, and $3{,}706$ ($33.8\%$) \textbf{unstable} structures. Sorting candidates into quartiles (Q1 to Q4) by their epistemic uncertainty reveals a strong negative correlation with robust dynamical stability (Spearman $r = -0.34, p < 10^{-300}$; Fig.~\ref{deep_fig_combined}a): $53\%$ of the structures in the lowest-uncertainty quartile (Q1) are classified as robustly stable ($32\%$ marginal, $15\%$ unstable), whereas the robustly stable fraction drops progressively to $39\%$ in Q2, $24\%$ in Q3, and only $11\%$ in the highest-uncertainty quartile (Q4, where $32\%$ are marginal and $57\%$ are unstable).

Crucially, checking the local second-order curvature through energy differential constraints ($\Phi_{ij}^{(k)} = \partial^2 E_G^{(k)} / \partial R_i^\alpha \partial R_j^\beta$) provides a physical mechanism to flag \textbf{high-suspicion candidates} that would otherwise pass deterministic MLIP screening as false positives. In crystalline materials, equilibrium forces often vanish by symmetry while smooth energy interpolation can produce a shallow, seemingly positive local curvature ($\omega_{\min}^{\mathrm{pt}} > -1.0$~THz). Whether harmonic force constants are evaluated via finite displacements, automatic differentiation, or derivative-free Hessian predictors such as HIP~\citep{Burger2026HIP}, relying on deterministic curvature alone cannot distinguish a deep, well-constrained potential well from an unreliable shallow curvature in sparsely sampled configuration space. Under EPIC's conservative differential perturbations, the dynamical matrix naturally amplifies local force uncertainties along soft lattice coordinates: when a candidate exhibits large imaginary-frequency excursions despite a non-negative point-predicted dispersion, it signals a high-suspicion regime (\textbf{marginal} class) where \emph{either the MLIP curvature prediction is locally inaccurate or the unrelaxed candidate crystal structure itself is physically unreasonable}, thereby prioritizing the structure for first-principles DFT verification. Figure~\ref{deep_fig_combined}b illustrates this diagnostic contrast by comparing $\mathrm{InSm}$ (left) and $\mathrm{BiSn}$ (right). For $\mathrm{InSm}$, the point-predicted dispersion is non-negative along the entire high-symmetry path ($\omega_{\min}^{\mathrm{pt}} = 0.004$~THz) with a modest mean band standard deviation of $\bar{\sigma}_{\omega} = 0.28$~THz ($\sigma_{\omega,\max} = 0.62$~THz at the $\mathrm{M}$-point acoustic branch, where $\omega = 0.32$~THz); all $32$ Monte Carlo realizations remain above the $-1.0$~THz threshold ($\omega_{\min,q05} = -0.89$~THz, $P_{\mathrm{stable}} = 100\%$), confirming its \textbf{robustly stable} assignment. By contrast, in $\mathrm{BiSn}$ (Fig.~\ref{deep_fig_combined}b, right), the deterministic point prediction also yields $\omega_{\min}^{\mathrm{pt}} = 0.004$~THz and would be accepted as dynamically stable by an uncalibrated MLIP screen. However, EPIC's energy-propagated differential check reveals severe curvature sensitivity across the zone-boundary acoustic and low-lying optical branches ($\bar{\sigma}_{\omega} = 1.35$~THz, $\sigma_{\omega,\max} = 2.68$~THz), driving $59.4\%$ of the Monte Carlo realizations below $-1.0$~THz with a 5th-percentile minimum of $\omega_{\min,q05} = -4.57$~THz ($P_{\mathrm{stable}} = 40.6\%$). An extended phonon dispersion gallery, summary table, and reciprocal-space perturbation-theory analysis explaining the $(2|\omega|)^{-1}$ spectral Jacobian amplification of soft modes are provided in Supplementary Note~10, Supplementary Figs.~S3 and S10, and Supplementary Table~S5.

Comparing energy epistemic uncertainty against the force-uncertainty prescreen score (Methods, Eq.~\eqref{eq:prescreen-score}) further demonstrates their complementary filtering value (Fig.~\ref{deep_fig_combined}c). Even among candidates with favorable (low) prescreen scores along the left edge of Figure~\ref{deep_fig_combined}c, structures with elevated epistemic uncertainty exhibit markedly depressed fractions of robust dynamical stability (dark red bins), whereas high stability fractions (blue and cyan bins) cluster tightly at low uncertainty percentiles along both axes.

At the chemical-element level, plotting the robustly stable fraction against mean epistemic uncertainty across all $81$ elements occurring in at least $15$ structures reveals a strong negative trend (Spearman $r = -0.63$, Fig.~\ref{deep_fig_combined}d), stratified sharply by bonding chemistry. Partitioning all $10{,}966$ candidates into five mutually exclusive chemical families (Supplementary Fig.~S4 and Supplementary Table~S6; see also the full $87$-element periodic-table and local coordination-entropy anatomy in Supplementary Note~7 and Supplementary Fig.~S7) shows that metallic and intermetallic systems (Other-Metals, $n = 9{,}166$) exhibit low mean phonon band uncertainty ($\bar{\sigma}_{\omega} = 0.59$~THz) and high robust stability ($3{,}393$ [$37.0\%$] robustly stable, $3{,}472$ [$37.9\%$] marginal, $2{,}301$ [$25.1\%$] unstable). By contrast, non-metal-containing families with strong ionic polarization, charge transfer, or light-mass vibrational modes exhibit substantially larger propagated phonon uncertainties and high fractions of unstable or marginal candidates: Oxides ($n = 1{,}064$: $2.1\%$ robustly stable, $11.6\%$ marginal, $86.4\%$ unstable; $\bar{\sigma}_{\omega} = 2.30$~THz), Halides ($n = 322$: $6.5\%$ robustly stable, $21.4\%$ marginal, $72.0\%$ unstable; $\bar{\sigma}_{\omega} = 1.41$~THz), Nitrides ($n = 272$: $9.6\%$ robustly stable, $29.8\%$ marginal, $60.7\%$ unstable; $\bar{\sigma}_{\omega} = 2.10$~THz), and Hydrides ($n = 142$: $11.3\%$ robustly stable, $26.1\%$ marginal, $62.7\%$ unstable; $\bar{\sigma}_{\omega} = 8.11$~THz). This chemistry-resolved audit demonstrates that EPIC naturally exposes the local-cutoff message-passing limitations of standard MLIPs in anionic and light-element bonding environments, preventing over-confident false-stable discoveries in high-throughput materials screening.

\section{Conclusion}\label{conclusion}

We have introduced EPIC (Energy-Propagated Intrinsic Confidence), an energy-centered evidential framework for uncertainty-aware machine-learned interatomic potentials. By placing a hierarchical evidential distribution over the potential energy, supervising atom-wise epistemic allocation with detached force residuals, and deriving point forces as exact gradients of the predicted mean energy, EPIC preserves strict conservative force fields while equipping both energies and forces with single-model uncertainty diagnostics.

Crucially, coupling the learned energy-level epistemic evidence with last-layer feature-Jacobian covariance enables EPIC to propagate scalar energy uncertainty directly into closed-form $3\times 3$ atomic force covariance matrices and conservative Monte Carlo force ensembles. Across all $79{,}020$ held-out MPtrj test structures ($7.48\times 10^6$ force components), raw evidential uncertainty provides a strong monotonic error-ranking signal (top-5\% error AUROCs of $0.891$ for energy and $0.883$ for forces), while two-parameter post-hoc variance recalibration restores quantitative 95\% interval coverage ($96.00\%$ for energy and $95.82\%$ for forces, calibrated force $\mathrm{NLL} = -0.6773$). When deployed as an active-learning gatekeeper, the combined EPIC energy-force trigger recalls $64.7\%$ of the top-5\% dangerous force-error configurations within a $10\%$ DFT review budget ($\mathrm{AUROC} = 0.912$), nearly doubling the recall of energy uncertainty alone.

Furthermore, propagating uncertainty across volumetric deformations, real-space force uncertainty ellipsoids, and harmonic phonon dispersions across $10{,}966$ candidate crystals exposes volumetric sampling biases in training databases, resolves sublattice-specific anionic uncertainty in complex oxides, and flags high-suspicion soft-mode instabilities where deterministic MLIP predictions appear falsely stable. As atomistic machine learning increasingly expands toward direct second-order curvature and Hessian prediction~\citep{Burger2026HIP}, our findings underscore the physical importance of enforcing exact energy-differential constraints ($E \to -\nabla E \to \nabla^2 E$) when assessing model reliability: only by propagating uncertainty self-consistently from the scalar potential energy surface to its first- and second-order spatial derivatives can hidden curvature distortions be systematically detected across thermodynamic, structural, and vibrational observables for autonomous materials discovery.

\section{Methods}\label{methods}

\noindent\textbf{Notation.} We distinguish the reference (DFT) total energy label $y$ (i.e., the reference potential energy up to the chosen DFT functional) from the model-predicted structure-level mean energy $\hat{E}(\mathbf{R}) \equiv \gamma_G(\mathbf{R})$, where the subscript $G$ denotes graph-level (structure-level) quantities and the subscript $i \in \{1,\dots,N\}$ denotes atom-level quantities. Unless otherwise stated, point forces and harmonic force constants are computed by differentiating the predicted mean energy $\gamma_G(\mathbf{R})$.

\subsection{Equivariant message passing backbone}\label{equivariant-message-passing-and-the-mace-architecture}

Modern MLIPs encode the local chemical environment of each atom through learned representations that respect the symmetries of three-dimensional Euclidean space. Among equivariant graph neural network potentials, the MACE architecture~\citep{Batatia2022MACE} achieves systematic improvability through a body-ordered expansion based on the Atomic Cluster Expansion (ACE) framework~\citep{Drautz2019ACE}, where the correlation order controls the many-body interactions captured by each message passing layer. Each atom $i$ is associated with a feature vector $\mathbf{h}_i$ that transforms under the irreducible representations of $O(3)$ up to order $L_{\max}$, and messages between neighboring atoms are constructed using tensor products of spherical harmonics and learnable radial functions. This body-ordered construction distinguishes MACE from equivariant architectures such as NequIP~\citep{Batzner2022} and Allegro~\citep{Musaelian2023Allegro}, which rely on iterative message passing to build up many-body correlations.

After $T$ interaction layers, the scalar invariant ($L=0$) component of the final node features is mapped to an atomic energy contribution $\hat{E}_i$ through a readout block. The structure-level predicted mean energy is obtained by summation:
\begin{equation}
\hat{E}(\mathbf{R}) = \sum_{i=1}^{N} \hat{E}_i(\mathbf{R}) + E_0 ,
\label{eq:mace-energy}
\end{equation}
where $E_0$ denotes per-species reference energy offsets. Atomic forces are computed exactly via automatic differentiation of the predicted mean energy:
\begin{equation}
\mathbf{F}_i = -\nabla_{\mathbf{R}_i}\hat{E}(\mathbf{R}) .
\label{eq:conservative-force}
\end{equation}
The conservative force construction in Eq.~\eqref{eq:conservative-force} is preserved throughout the EPIC formulation.

\subsection{Evidential deep learning preliminaries}\label{evidential-deep-learning}

Evidential Deep Learning (EDL), introduced by Amini et al.~\citep{Amini2020DER}, offers a principled single-pass approach to uncertainty quantification by placing a higher-order prior over the parameters of the predictive Gaussian likelihood $\mathcal{N}(y \mid \mu, \sigma^2)$. Treating both $\mu$ and $\sigma^2$ as latent random variables, EDL assigns them a conjugate Normal-Inverse-Gamma (NIG) prior through a conditional-marginal decomposition:
\begin{equation}
\begin{aligned}
p(\mu, \sigma^2)
&= p(\mu \mid \sigma^2)p(\sigma^2) \\
&= \mathcal{N}\!\left(\mu \mid \gamma_G, \frac{\sigma^2}{\nu_G}\right)
   \operatorname{Inv\text{-}Gamma}(\sigma^2 \mid \alpha_G, \beta_G).
\end{aligned}
\label{eq:nig-prior}
\end{equation}
The neural network outputs the four structure-level hyperparameters $(\gamma_G, \nu_G, \alpha_G, \beta_G)$, where $\gamma_G$ is the expected predictive mean, $\nu_G > 0$ is the virtual-observation precision scaling factor for $\mu \mid \sigma^2$, and $\alpha_G > 1, \beta_G > 0$ are the shape and scale parameters of the inverse-gamma prior on $\sigma^2$.

Marginalizing over $\sigma^2$ yields a Student-$t$ marginal distribution with $2\alpha_G$ degrees of freedom. From the NIG moments, the structure-level aleatoric, epistemic, and total predictive variances are disentangled in closed form:
\begin{equation}
\sigma_{\mathrm{al},G}^2 = \mathbb{E}[\sigma^2]
= \frac{\beta_G}{\alpha_G - 1},
\quad
\sigma_{\mathrm{ep},G}^2 = \operatorname{Var}[\mu]
= \frac{\beta_G}{\nu_G(\alpha_G - 1)},
\quad
\sigma_{\mathrm{tot},G}^2 = \sigma_{\mathrm{al},G}^2 + \sigma_{\mathrm{ep},G}^2 .
\label{eq:uncertainty-decomposition}
\end{equation}
Compared to deep ensembles~\citep{Lakshminarayanan2017DeepEnsembles}, which require training and evaluating $M$ independent models, EDL estimates both uncertainty components in a single forward pass, though its raw variance scale can exhibit under- or over-confidence requiring post-hoc calibration~\citep{Bengs2022Pitfalls}.

\subsection{Energy-force consistency and EPIC formulation}\label{the-energy-force-consistency-problem}

The physical relation $\mathbf{F}_i = -\nabla_{\mathbf{R}_i} E(\mathbf{R})$ distinguishes two notions of consistency: (i) mean-level conservation, where point forces are exact gradients of a scalar potential $\gamma_G(\mathbf{R})$, and (ii) uncertainty-level propagation, where energy and force uncertainties stem from a unified potential energy surrogate. Recent force-level evidential models such as eIP~\citep{Xu2026EvidentialIP} and e$^2$IP~\citep{Wang2026EquivariantEDLIP} parameterize NIG or Normal-Inverse-Wishart priors directly on force components, providing valuable local force diagnostics but leaving the potential energy uncertainty undefined without additional integrability assumptions. Conversely, EPIC places the evidential distribution directly on the potential energy and propagates uncertainty to forces and second-derivative observables through coordinate differentiation of the learned feature representation.

\subsection{Energy-level evidential formulation and hierarchical allocation}\label{energy-level-evidential-formulation}

We implement the EPIC readout module on top of the MACE invariant node features $\mathbf{h}_i$ as illustrated in Figure~\ref{fig:epic-architecture} (an expanded six-stage mathematical flowchart is also provided in Supplementary Fig.~S5).

\begin{figure}[!htbp]
\centering
\includegraphics[width=\linewidth]{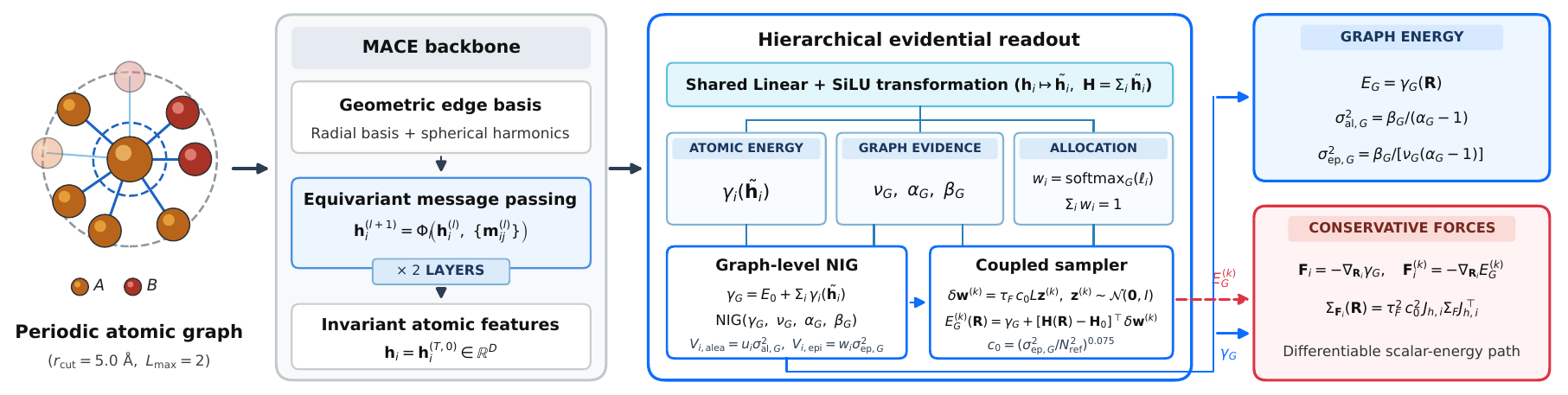}
\caption{Architecture of the MACE+EPIC model. Equivariant MACE message passing produces invariant atomic features $\mathbf{x}_i = \mathbf{h}_i^{(T,0)}$ that pass through a shared linear transformation and SiLU activation. An atomic energy head outputs mean-energy contributions $E_i \equiv \gamma_i$, while mean-pooled features feed the structure-level NIG evidence heads $(\nu_G, \alpha_G, \beta_G)$ and an atom-wise softmax allocation head produces epistemic weights $w_i$ ($\sum_i w_i = 1$). Differentiating the structure-level mean energy $\gamma_G$ yields exact conservative point forces $\mathbf{F}_i = -\nabla_i E_G$, while the coupled energy-propagated feature sampler constructs perturbed conservative energy surfaces $E_G^{(k)}(\mathbf{R})$ and closed-form $3\times 3$ atomic force covariance matrices $\Sigma_{\mathbf{F}_i} = \tau_F^2\,c_0^2\,J_{h,i}\Sigma_F J_{h,i}^{\top}$.}
\label{fig:epic-architecture}
\end{figure}

Specifically, the scalar invariant features $\mathbf{h}_i$ from the final MACE interaction layer are mapped into shared hidden representations and mean-pooled across the structure:
\begin{equation}
\tilde{\mathbf{h}}_i
=
\phi\!\left(W_{\mathrm{shared}}\mathbf{h}_i\right),
\qquad
\bar{\mathbf{h}}
=
\frac{1}{N}\sum_{i=1}^{N}\tilde{\mathbf{h}}_i,
\label{eq:shared-hidden}
\end{equation}
where $\phi$ denotes the second-moment-normalized SiLU activation. The EPIC readout heads then predict
\begin{equation}
\begin{aligned}
\gamma_i
&= W_{\gamma}\tilde{\mathbf{h}}_i, \\
\nu_G
&= \operatorname{softplus}(W_{\nu}\bar{\mathbf{h}})+\epsilon, \\
\alpha_G
&= \operatorname{softplus}(W_{\alpha}\bar{\mathbf{h}})+1+\epsilon, \\
\beta_G
&= \operatorname{softplus}(W_{\beta}\bar{\mathbf{h}})+\epsilon,
\end{aligned}
\label{eq:epic-heads}
\end{equation}
with $\epsilon=10^{-6}$, guaranteeing $\nu_G>0$, $\alpha_G>1$, and $\beta_G>0$. The structure-level predicted mean energy is
\begin{equation}
\hat{E}(\mathbf{R})\equiv\gamma_G(\mathbf{R})
=
\sum_{i=1}^{N}\gamma_i(\mathbf{R})+E_0,
\label{eq:nig-aggregation}
\end{equation}
and deterministic conservative forces follow via $\mathbf{F}_i = -\nabla_{\mathbf{R}_i}\gamma_G(\mathbf{R})$.

To resolve site-specific uncertainty while preserving exact additive consistency with the graph-level variance, an allocation head maps $\tilde{\mathbf{h}}_i$ to logits $\ell_i$, which pass through a structure-wise softmax with a positive floor and renormalization to yield epistemic allocation weights $w_i = \operatorname{softmax}_G(\ell_i) > 0$ satisfying $\sum_{i=1}^N w_i = 1$. Using uniform aleatoric weights $u_i = 1/N$, the atom-wise NIG parameters and allocated variances are defined as
\begin{equation}
\alpha_i=\alpha_G,
\qquad
\beta_i=u_i\beta_G,
\qquad
\nu_i=\nu_G\frac{u_i}{w_i},
\label{eq:atomic-nig-allocation}
\end{equation}
\begin{equation}
V_{i,\mathrm{alea}}
=
u_i\frac{\beta_G}{\alpha_G-1} = u_i\,\sigma_{\mathrm{al},G}^2,
\qquad
V_{i,\mathrm{epi}}
=
w_i\frac{\beta_G}{\nu_G(\alpha_G-1)} = w_i\,\sigma_{\mathrm{ep},G}^2,
\label{eq:atomic-variance-allocation}
\end{equation}
such that $\sum_{i=1}^N V_{i,\mathrm{alea}} = \sigma_{\mathrm{al},G}^2$ and $\sum_{i=1}^N V_{i,\mathrm{epi}} = \sigma_{\mathrm{ep},G}^2$ hold identically.

\subsection{Training objective and regularization}\label{training-objective-and-regularization}

The structure-level NIG parameters $(\gamma_G,\nu_G,\alpha_G,\beta_G)$ are trained using the exact Student-$t$ marginal negative log-likelihood of the Gaussian--NIG hierarchy (derivation in Supplementary Note~1):
\begin{equation}
\begin{aligned}
\mathcal{L}_{\mathrm{NLL}}
={}&
\frac{1}{2}\log\left(\frac{\pi}{\nu_G}\right)
-\alpha_G\log\Omega_G \\
&+
\left(\alpha_G+\frac{1}{2}\right)
\log\left[\nu_G(y-\gamma_G)^2+\Omega_G\right] \\
&+
\log\Gamma(\alpha_G)
-\log\Gamma\left(\alpha_G+\frac{1}{2}\right),
\end{aligned}
\label{eq:nll-loss}
\end{equation}
where $\Omega_G=2\beta_G(1+\nu_G)$ and $y$ is the reference DFT total energy. To penalize spurious evidence on high-error configurations, we include the evidential regularization term~\citep{Amini2020DER}
\begin{equation}
\mathcal{L}_{\mathrm{reg}}
=
|y-\gamma_G|(2\nu_G+\alpha_G),
\label{eq:evidential-regularization}
\end{equation}
whose system-size scaling characteristics are analyzed in Supplementary Note~2 (Table~S1). To guide the atomic epistemic allocation weights $w_i$ toward locally uncertain atomic environments without corrupting the conservative energy and force backbone, we supervise $w_i$ using detached squared force residuals $e_i^{\mathrm{detach}} = \|\mathbf{F}_i^{\mathrm{detach}} - \mathbf{F}_i^{\mathrm{DFT}}\|_2^2$ normalized within each structure ($\tilde{e}_i = e_i^{\mathrm{detach}} / \sum_{j=1}^N e_j^{\mathrm{detach}}$):
\begin{equation}
\mathcal{L}_{\mathrm{alloc}}
=
-\sum_{i=1}^{N} \tilde{e}_i \log w_i ,
\qquad
\mathbf{F}_i^{\mathrm{detach}} = \operatorname{stop\_grad}\!\left(-\nabla_{\mathbf{R}_i}\gamma_G(\mathbf{R})\right).
\label{eq:alloc-loss}
\end{equation}
The complete training objective is
\begin{equation}
\mathcal{L}
=
\left\langle
\mathcal{L}_{\mathrm{NLL}}
+
\lambda(t)\mathcal{L}_{\mathrm{reg}}
+
\lambda_{\mathrm{alloc}}\mathcal{L}_{\mathrm{alloc}}
\right\rangle_{\mathcal{B}}
+
w_f\mathcal{L}_{\mathrm{F}}^{\mathrm{cond.\,Huber}},
\label{eq:total-loss}
\end{equation}
where $\langle\cdot\rangle_{\mathcal{B}}$ denotes minibatch averaging and $\mathcal{L}_{\mathrm{F}}^{\mathrm{cond.\,Huber}}$ is a conditional Huber loss on $\mathbf{F}_i = -\nabla_{\mathbf{R}_i}\gamma_G(\mathbf{R})$ with base threshold $0.01$~eV/\AA{}, $w_f = 2000$, $\lambda_{\mathrm{alloc}} = 0.1$, and $\lambda(t)$ linearly annealed to $10^{-3}$ over the first five epochs (full training hyperparameters in Supplementary Note~3, Table~S2).

\subsection{Energy-propagated force uncertainty and downstream observable propagation}\label{uncertainty-propagation-to-downstream-observables}

\paragraph{Energy-propagated force covariance and conservative MC sampling.}
While mean forces are deterministically derived from $\gamma_G(\mathbf{R})$, EPIC constructs both closed-form $3\times 3$ atomic force covariance matrices and conservative Monte Carlo force realizations by coupling the structure-level epistemic energy variance $\sigma_{\mathrm{ep},G}^2(\mathbf{R})$ to the coordinate Jacobians of the pooled shared representation $\mathbf{H}(\mathbf{R}) = \sum_{i=1}^N \tilde{\mathbf{h}}_i(\mathbf{R}) \in \mathbb{R}^D$. Let $J_{h,i}(\mathbf{R}) = \partial\mathbf{H}(\mathbf{R})/\partial\mathbf{R}_i \in \mathbb{R}^{3\times D}$ denote the coordinate Jacobian of the pooled features on atom $i$, and let
\begin{equation}
G_F = \sum_{c\in\mathcal{D}_{\mathrm{cal}}}\sum_{i=1}^{N_c} J_{h,i}(\mathbf{R}_c)^{\top}J_{h,i}(\mathbf{R}_c),
\qquad
\Sigma_F = \left(G_F + \lambda_{\mathrm{cov}} I_D\right)^{-1} = LL^{\top} \in \mathbb{R}^{D\times D}
\label{eq:sigma-f-gram}
\end{equation}
denote the regularized inverse Gram covariance matrix of the force-feature Jacobians estimated on the disjoint calibration split $\mathcal{D}_{\mathrm{cal}}$ with $\lambda_{\mathrm{cov}} = 10^{-3}\operatorname{tr}(G_F)/D$. For a structure with reference configuration $\mathbf{R}_{\mathrm{ref}}$ of $N_{\mathrm{ref}}$ atoms, the damped energy-evidence reference scale is defined as
\begin{equation}
c_0(\mathbf{R}_{\mathrm{ref}})
=
\left(
\frac{\sigma_{\mathrm{ep},G}^2(\mathbf{R}_{\mathrm{ref}})}{N_{\mathrm{ref}}^2}
\right)^{\eta},
\qquad
\eta = 0.075 ,
\label{eq:c0-scale}
\end{equation}
corresponding to a power-law damping exponent of $2\eta = 0.15$ on the per-atom epistemic variance that prevents extreme heavy-tailed NIG fluctuations from destabilizing force covariance estimates.

For $k = 1,\dots,K$ Monte Carlo realizations, we sample standard normal latent vectors $\mathbf{z}^{(k)} \sim \mathcal{N}(\mathbf{0}, I_D)$ and construct perturbed scalar potential energy surfaces:
\begin{equation}
\delta\mathbf{w}^{(k)}
=
\tau_F\,c_0 L\mathbf{z}^{(k)},
\qquad
E_G^{(k)}(\mathbf{R})
=
\gamma_G(\mathbf{R})
+
\left[\mathbf{H}(\mathbf{R})-\mathbf{H}_0\right]^{\top}\delta\mathbf{w}^{(k)},
\label{eq:mc-energy-sampling}
\end{equation}
where $\mathbf{H}_0 = \mathbf{H}(\mathbf{R}_{\mathrm{ref}})$ is anchored at the undisplaced reference configuration and $\tau_F$ is the multiplicative force standard-deviation calibration factor (with variance scale $a \equiv \tau_F^2$). Differentiating $E_G^{(k)}(\mathbf{R})$ with respect to $\mathbf{R}_i$ with $\delta\mathbf{w}^{(k)}$, $c_0$, and $\mathbf{H}_0$ held fixed across displaced geometries yields conservative force realizations
\begin{equation}
\mathbf{F}_i^{(k)}(\mathbf{R})
=
-\nabla_{\mathbf{R}_i}E_G^{(k)}(\mathbf{R})
=
\mathbf{F}_i(\mathbf{R}) - J_{h,i}(\mathbf{R})\,\delta\mathbf{w}^{(k)}.
\label{eq:mc-force}
\end{equation}
Taking the analytical expectation over $\mathbf{z}^{(k)} \sim \mathcal{N}(\mathbf{0},I_D)$ yields the exact closed-form $3\times 3$ atomic force covariance tensor:
\begin{equation}
\Sigma_{\mathbf{F}_i}(\mathbf{R})
=
\operatorname{Cov}\!\left[\mathbf{F}_i^{(k)}(\mathbf{R})\right]
=
\tau_F^2\,c_0^2\,J_{h,i}(\mathbf{R})\,\Sigma_F\,J_{h,i}(\mathbf{R})^{\top},
\label{eq:closed-form-force-cov}
\end{equation}
and the structure-level root-mean-square force uncertainty is $\mathrm{RMS}(\sigma_F) = \sqrt{\frac{1}{3N}\sum_{i=1}^N \operatorname{tr}(\Sigma_{\mathbf{F}_i}(\mathbf{R}))}$. For quantitative Gaussian log-likelihood (NLL) and interval coverage evaluations across the test set, the marginal variance along Cartesian axis $\alpha \in \{x,y,z\}$ incorporates an additive noise floor $\sigma_{0,F}^2$ fitted alongside $\tau_F$ on the disjoint calibration split:
\begin{equation}
\left(\sigma_{F,i\alpha}^{\mathrm{cal}}(\mathbf{R})\right)^2
=
\left(\Sigma_{\mathbf{F}_i}(\mathbf{R})\right)_{\alpha\alpha} + \sigma_{0,F}^2
=
\tau_F^2\,c_0^2\left(J_{h,i}(\mathbf{R})\,\Sigma_F\,J_{h,i}(\mathbf{R})^{\top}\right)_{\alpha\alpha} + \sigma_{0,F}^2,
\label{eq:force-affine-cal}
\end{equation}
with fitted parameters $\tau_F = 48.2697$ ($a = \tau_F^2 = 2329.97$) and $\sigma_{0,F} = 0.003374$~eV/\AA{} (and analogously $(\sigma_{E}^{\mathrm{cal}})^2 = \tau_E^2(\sigma_{\mathrm{ep},G}/N)^2 + \sigma_{0,E}^2$ with $\tau_E = 0.7278$ and $\sigma_{0,E} = 0.0365$~eV/atom for per-atom energy). When propagating conservative force perturbations to harmonic phonon dynamical matrices via Eq.~\eqref{eq:mc-force}, only the multiplicative scale $\tau_F$ enters the coordinate derivatives while the additive constant $\sigma_{0,F}^2$ drops out under differentiation.

\paragraph{Structural uncertainty ellipsoids.}
From the $3\times 3$ atomic force covariance matrix $\Sigma_{\mathbf{F}_i}(\mathbf{R})$ in Eq.~\eqref{eq:closed-form-force-cov}, eigendecomposition $\Sigma_{\mathbf{F}_i} = U_i\operatorname{diag}(\lambda_{i,1},\lambda_{i,2},\lambda_{i,3})U_i^{\top}$ defines the principal standard-deviation semi-axes $\sqrt{\lambda_{i,m}}$ ($m=1,2,3$) and orientation vectors, which are exported in crystallographic CIF \texttt{\_atom\_site\_aniso\_U\_*} format and rendered in VESTA (normalized per crystal for visual clarity of site-to-site directional anisotropy, with exact physical magnitudes $\mathrm{RMS}(\sigma_F)$ and $\max_i\sqrt{\lambda_{i,3}}$ in eV/\AA{} reported in each panel header).

\paragraph{Equation of state.}
For volumetric equation-of-state evaluations, each crystal structure is isotropically scaled across $16$ volume points spanning $V/V_0 \in [0.85, 1.15]$, and the predicted mean energy $\gamma_G(V)/N$, total uncertainty $\sigma_{\mathrm{tot},G}(V)/N$, and epistemic uncertainty $\sigma_{\mathrm{ep},G}(V)/N$ are evaluated at each volume to compare the physical equilibrium volume $V_0 = \arg\min_V \gamma_G(V)$ against the minimum-epistemic-uncertainty volume $V_{\min}^{\mathrm{epist}} = \arg\min_V \sigma_{\mathrm{ep},G}(V)$.

\paragraph{Prescreen score for candidate prioritization.}
To rapidly prioritize candidate structures prior to expensive supercell phonon evaluations, we define a normalized prescreen score
\begin{equation}
S_{\mathrm{pre}}
=
\sum_j w_j x_j',
\qquad
x_j'
=
\frac{x_j - \min(x_j)}{\max(x_j) - \min(x_j)} ,
\label{eq:prescreen-score}
\end{equation}
where $x_j'$ denotes min-max (or percentile) normalization across the candidate pool. In the two-dimensional stability map (Fig.~\ref{deep_fig_combined}c), $S_{\mathrm{pre}}$ specifically evaluates the structure-level energy-propagated force uncertainty percentile $\mathrm{RMS}(\sigma_F)$ along the horizontal axis against the per-atom energy epistemic uncertainty percentile $\sigma_{\mathrm{ep},G}/N$ along the vertical axis, isolating the complementary screening power of force- and energy-level confidence; more generally, Eq.~\eqref{eq:prescreen-score} also supports multi-criteria pre-filtering combining epistemic uncertainty, total uncertainty, maximum atomic force magnitude, and unit-cell size.

\paragraph{Phonon dispersions and dynamical stability classification.}
Harmonic phonon dispersions were computed using the finite-displacement supercell method implemented in \texttt{Phonopy}~\citep{TOGO20151} with a $2\times 2\times 2$ supercell expansion and a finite atomic displacement amplitude of $0.01$~\AA{}. For each candidate crystal, the deterministic harmonic force constants $\Phi_{ij}^{\alpha\beta} = \partial^2 \gamma_G / \partial R_i^\alpha \partial R_j^\beta$ and $K = 32$ perturbed force-constant tensors $\Phi_{ij}^{\alpha\beta,(k)} = \partial^2 E_G^{(k)} / \partial R_i^\alpha \partial R_j^\beta$ were assembled by evaluating the deterministic forces $\mathbf{F}(\mathbf{R})$ and conservative Monte Carlo force realizations $\mathbf{F}^{(k)}(\mathbf{R})$ (Eq.~\eqref{eq:mc-force}, with $c_0$, $\mathbf{H}_0$, and $\delta\mathbf{w}^{(k)}$ held fixed across all displaced supercells) and subtracting the corresponding undisplaced supercell reference forces prior to fitting the force constants.

Along the standardized \texttt{Seekpath} high-symmetry Brillouin-zone path, we extracted the baseline minimum phonon frequency $\omega_{\min}^{\mathrm{pt}}$ from the deterministic force constants and the sample minimum frequencies $\{\omega_{\min}^{(k)}\}_{k=1}^{32}$ from the $K=32$ Monte Carlo realizations. Using an empirical screening tolerance of $\omega_{\mathrm{tol}} = -1.0$~THz to accommodate minor numerical acoustic-sum-rule and finite-supercell discretization residuals, let $P_{\mathrm{stable}} = \frac{1}{K}\sum_{k=1}^K \mathbb{I}(\omega_{\min}^{(k)} \ge \omega_{\mathrm{tol}})$ denote the stable MC fraction and $\omega_{\min,q05}$ denote the 5th percentile of $\{\omega_{\min}^{(k)}\}_{k=1}^{32}$. Each candidate was assigned to one of three dynamical stability classes:
\begin{itemize}
    \item \textbf{Robustly stable:} $\omega_{\min}^{\mathrm{pt}} > \omega_{\mathrm{tol}}$, $P_{\mathrm{stable}} \ge 0.95$, and $\omega_{\min,q05} > \omega_{\mathrm{tol}}$.
    \item \textbf{Marginal (high-suspicion soft-mode candidates):} $\omega_{\min}^{\mathrm{pt}} > \omega_{\mathrm{tol}}$, but $P_{\mathrm{stable}} < 0.95$ or $\omega_{\min,q05} \le \omega_{\mathrm{tol}}$, indicating that energy-propagated force uncertainty induces imaginary-frequency excursions below $-1.0$~THz and flagging the structure for DFT verification.
    \item \textbf{Unstable:} $\omega_{\min}^{\mathrm{pt}} \le \omega_{\mathrm{tol}}$, irrespective of the Monte Carlo uncertainty.
\end{itemize}

\section*{Code availability}
The code used to generate the results in this study is openly available at \url{https://github.com/SuthPhy2Ai/EPIC}.

\section*{Acknowledgements}
This work is supported by the Advanced Materials-National Science and Technology Major Project (2025ZD0618802) and the Shanghai Technical Service Center of Science and Engineering Computing, Shanghai University.

\section*{Author contributions}
Tianhao Su and Musen Li contributed equally to this work. Tianhao Su: Conceptualization, Methodology, Software, Investigation, Data curation, Visualization, Writing-original draft. Musen Li: Methodology, Software, Formal analysis, Validation, Data curation, Writing-review \& editing. Shunbo Hu: Conceptualization, Methodology, Formal analysis, Investigation, Supervision, Funding acquisition, Writing-original draft, Writing-review \& editing. Tong-Yi Zhang: Supervision, Project administration, Resources, Writing-review \& editing.

\section*{Competing interests}
The authors declare no competing interests.

\section*{Additional information}
Supplementary Information is available as a separate file.

\bibliography{references}

\end{document}